**Research article**

# DNA insertion mutations can be predicted by a periodic probability function

Tatsuaki Tsuruyama[1]

*[1]Department of Pathology, Graduate School of Medicine, Kyoto University, Yoshida Konoe-cho, Sakyo-ku, Kyoto 606-8501, Japan*

**Running title:** Probability prediction of mutation

**Correspondence to:** Tatsuaki Tsuruyama

Kyoto University, Yoshida-Konoe-cho, Sakyo-ku, Kyoto 606-8501, Japan

Tel.: +81-75-751-3488; Fax: +81-75-761-9591

E-mail: tsuruyam@kuhp.kyoto-u.ac.jp

**Abstract**

It is generally difficult to predict the positions of mutations in genomic DNA at the nucleotide level. Retroviral DNA insertion is one mode of mutation, resulting in host infections that are difficult to treat. This mutation process involves the integration of retroviral DNA into the host-infected cellular genomic DNA following the interaction between host DNA and a pre-integration complex consisting of retroviral DNA and integrase. Here, we report that retroviral insertion sites around a hotspot within the *Zfp521* and *N-myc* genes can be predicted by a periodic function that is deduced using the diffraction lattice model. In conclusion, the mutagenesis process is described by a biophysical model for DNA–DNA interactions.

**Text**

**Introduction**

Extensive research has examined retroviral insertions to further our understanding of DNA mutations. Retrovirus-related diseases, including leukemia/lymphoma and AIDS, develop after retroviral genome insertion into the genomic DNA of the infected host cell. Retroviral DNA insertion is one of the modes of insertional mutation. After reverse-transcription of the retroviral genomic RNA into DNA, the retroviral DNA forms a pre-insertion complex (PIC) with the integrase enzyme, which catalyzes the insertion reaction. The PIC crosses the nuclear membrane via a nuclear pore complex to access the host cell DNA. The terminal -OH groups of MLV DNA attack the host DNA by introducing a nick in a staggered manner within a target four-base region. The reaction is completed at the stage of the duplication of the four bases (**Figure 1a**). This interaction between the PIC and host DNA has been believed to be a random process, independent of the primary DNA sequence of the host; however, MLV DNA became part of the mouse DNA, and therefore was classified as an insertion mutation. a preference for palindromic sequence motifs is expected, to some degree (1, 9, 10).

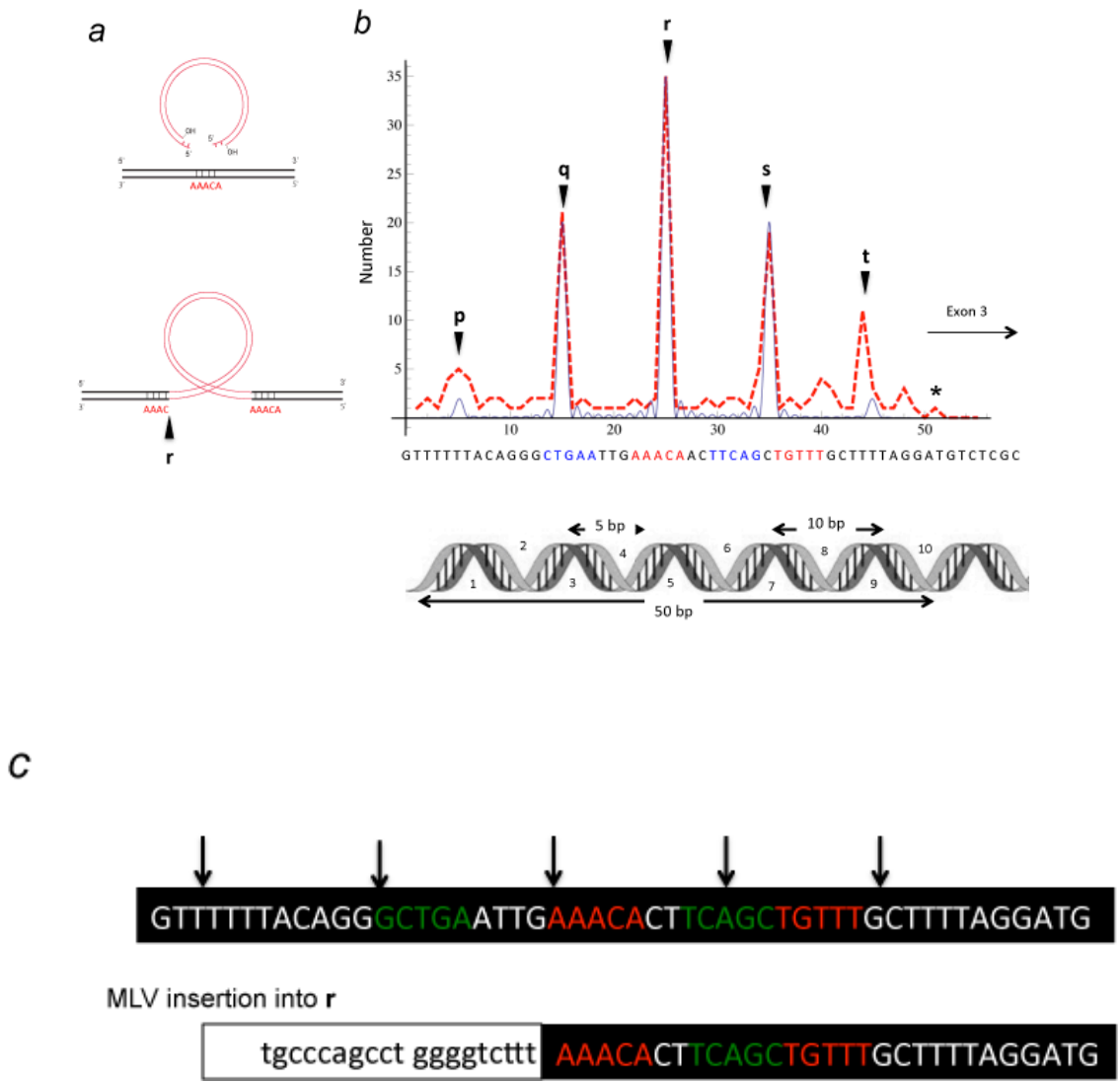

**Fig. 1. Probability function predicting MLV insertion in *Zfp521***
(***a***) The most frequent insertion pattern of the MLV genome *r* in the right plot. The red letters AAAC represent the target oligonucleotides that are duplicated after the insertion. (***b***) Numbers of mice with individual MLV genome insertions in LN cells are shown by the red dotted line (total number of mice *n* = 136, total insertions = 169); the wedge sites labeled **p, q, r, s,** and **t** indicate spots of frequent MLV genome insertion at ten-base intervals and the dominant peaks of light diffraction. The insertion into the segment occurred one or two times in individual mice over their lifetime based on PCR analyses (total insertions = 169). The curve illustrates the function of light diffraction pattern resulting from a 10-slit grating (Eq. 1, See Supplementary Material). The bases at the bottom indicate the *Zfp521* target segment sequence. Red and blue letters represent palindromic motifs that alternate. A double-stranded DNA structure model is shown at the bottom. Underlining represents the four oligonucleotides duplicated by the insertion in the host cell genome when the insertion occurs at **p, q, r, s,** and **t**. The **q** and **r** positions are identical to previously reported positions (1). (*c*) The junctional sequence of MLV and host DNA. The case of MLV insertion into Zfp521 at the **r** position is shown. Green and red letters represent pairs of alternating palindromes. The sequences in the same colors are palindromes.

The retroviral DNA–host DNA interaction favors specific host DNA structures that depends on primary sequence (2-7) A previous study has shown that insertion sites in HIV-1 are distributed with a 10-bp period on the nucleosome surface, specifically on the outward-facing major grooves in chromatin (2, 8). However, it has not been possible to predict the probability of insertion mutations at specific nucleotides.

To study murine leukemia retrovirus (MLV) insertions, an inbred mouse strain (SL/Kh) has been developed in RIKEN (see materials). The SL/Kh mice genetically harbor an integrated MLV genome (AKV1) as a part of their own genome, which is transmitted through the germ line from the ancestral strain AKR (9). At birth, SL/Kh mice spontaneously develop an active systemic of MLV particle, which originates from AKV1. MLV particle selectively infects B cells via tropism that is determined by the affinity between a viral coat protein and a cell-surface protein on host B cells. Finally, AKV1 retroviral DNA is integrated into the B-cell genome. We found that such re-insertions of the AKV1 genome were concentrated in the zinc finger protein 521 gene (*Zfp521*). Greater than 95% of all SL/Kh mice had an AKV1 insertion in *Zfp521* in B-cells at 3 weeks of age, but lacked clinical symptoms, followed by clonal lymphocytes. This gene probably contributes to chondrocyte development (10), in addition to B cell development, which requires an abnormal chimeric gene (11).

**Methods**

**Mouse strain.** SL/Kh strains are bred by the RIKEN BioResource Center (http://en.brc.riken.jp/) and are widely available (12). All mice used in this study were handled in strict accordance with the guidelines for good animal practice, as defined by the relevant national and local animal welfare bodies, and all animal work was preapproved by the Kyoto University Ethics Committee for Animal Experiments (See Supporting Material).

**PCR for identification of MLV integration site.**

To directly identify MLV insertion into the second intron of *Zfp521*, a forward primer (1F: 5'-CGGCCACGAGGAAGTGTAG-3') in *Zfp521*, a reverse primer in MLV (1R: 5'-TACAGAAGCGAGAAGCGAGC-3'), and a nested reverse PCR primer (2R 5'-AGTGCTTGACCACAGATATCC-3') (8233-8252) were used. To identify the insertion into the first exon of *Zfp521*, we used primers 1F: 5'-ACCCCACACTCGCGCCAGTCC-3' and 1R: 5'-CTGCAAGG TCTTGGACTGCG-3', along with a nested reverse PCR primer, 2F: 5'-TCCGATAGACTGAGTCGCC-3'. To identify AKV1 insertion into *N-myc*, a forward primer (1F: 5'- GCTGGTGAAGAACGAGAAGG-3') and a nested forward PCR primer (2F: 5'- AAAGAAGATCGAACACGCTC-3') in *N-myc*, a reverse primer in AKV1 (1R: 5'-TACAGAAGCGAGAAGCGAGC-3'), and a nested reverse PCR primer (2R 5'-AGTGCTTGACCACAGATATCC-3') were used. To identify AKV1 insertion into *N-myc*, a forward primer (1F: 5'- GCTGGTGAAGAACGAGAAGG-3') and a nested forward PCR primer (2F: 5'- AAAGAAGATCGAACACGCTC-3') in *N-myc*, a reverse primer in AKV1 (1R: 5'-TACAGAAGCGAGAAGCGAGC-3'), and a nested reverse PCR primer (2R 5'-AGTGCTTGACCACAGATATCC-3') were used. The virus-host junctions were amplified in a 50 μl reaction mixture containing 2.5 m*M* dNTP, 10 pmol/ml primer, and 0.25 unit of *Taq* polymerase (Expand Long Template PCR System) (Roche Diagnostics, Mannheim, Germany).

**Results**

**Probability function predicting insertion mutations in *Zfp521***

The insertion sites were identified using inverse PCR (see Methods). The target gene *Zfp521* carries an alternating palindromic sequence (13, 14) that is frequently targeted by MLV insertions. MLV DNA forms a complex with integrase, which catalyzes the insertion reaction, yielding a PIC.

Each insertion occurred once or twice during the lifetime of the mouse within a 50-bp segment located in the second intron to the third exon of *Zfp521*. Most commonly, the insertion occurred at the ":" site in the sequence "CTGAATTG*AAAC: A*ACTTCAGC*TGTTT*," where the pair of underlined sequences and the pair of sequences in italics are palindromic. The insertions did not occur randomly; rather, they frequently occurred at approximately 10-bp intervals and were symmetrically distributed on either side of the most frequent hot spot, **r**, within the shown 50-bp segment (**Figure 1b**) (13, 14). The junction from the MLV to the host mouse DNA sequence is shown in **Figure 1c**. The data for plotting is given in supporting materials.

**Probability function predicting insertions in *N-myc***

A similar distribution to the pattern of insertions in *Zfp521* was observed for insertions in the third exon and 3'-flanking sequence of *N-myc* in the genome of lymph node (LN) cells ($n = 83$) (**Fig. 2**).

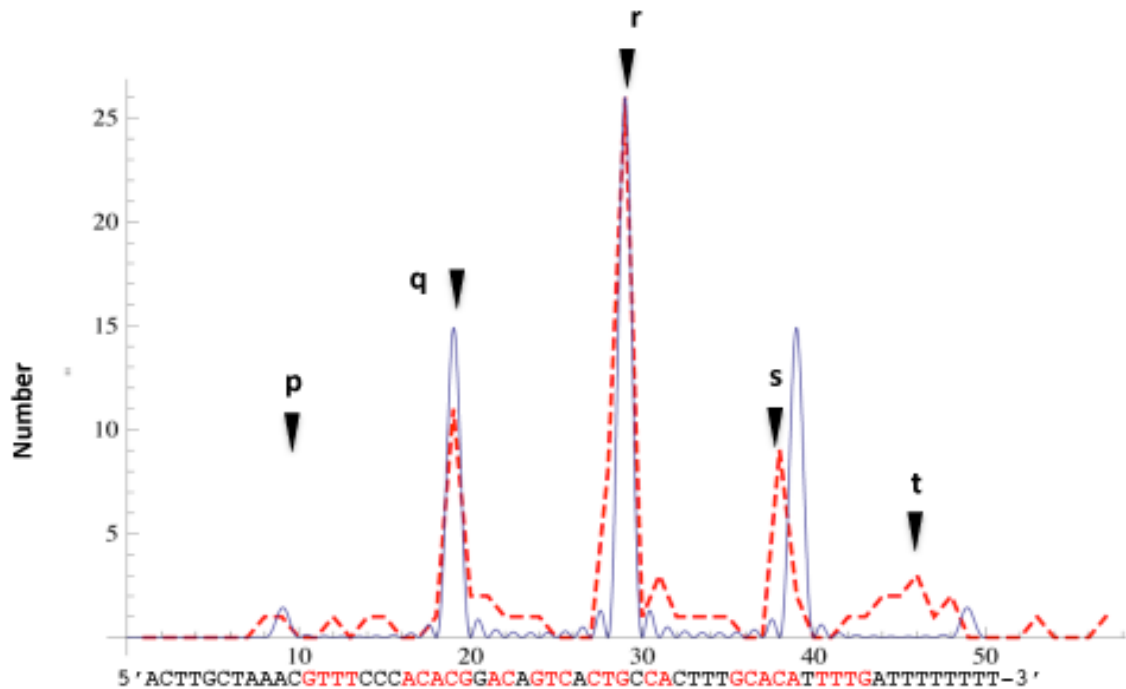

**Fig. 2. Probability function predicting MLV insertions within *N-myc***
The numbers of mice with individual MLV genome insertions at individual nucleotides in the *N-myc* (No. 6300-, M12731) gene in LN cells are shown by the red dotted line (total number of mice *n* = 136, total insertions = 83); the wedge sites labeled **p, q, r, s,** and **t** indicate spots of frequent MLV genome insertion. The curve illustrates the function for insertion sites in the *N-myc* segments. The **q, r, s,** and **t** positions are identical to the positions previously reported. Red letters represent a palindromic motif. Underlining represents the nucleotides duplicated by the insertion when the insertion occurs at **p, q, r, s,** and **t**.

Most commonly, the insertion occurred at the ":" site (**r** position in **Fig. 2**) in the sequence "*ACGTTTC*CCA<u>CA</u>CACG<u>GACA</u>: <u>GTCA</u>CT<u>G</u>C*CACTTTGCA*C," where the pair of underlined sequences are palindromic sequences and the pair of sequences in italics are mirror symmetric sequences(15-18). As a result of the insertion, MLV DNA became part of the mouse DNA, as aforementioned. The data for plotting is given in supporting materials.

**A prediction model of the probability function**

We noted that the patterns of insertion sites and frequency within *Zfp521* and *N-myc* were similar to the diffraction pattern after a hypothetical wave passes through a gate. This idea is according to Arndt et al. (19-22), who described the diffraction pattern of the macromolecule $C_{60}$. The detailed calculation is descried in the supplemental information. The grating contains ten slits separated by a slit interval that is five times greater than the width of a single slit of which proportion is close to the actual proportion of the DNA structure and validates following model and calculation. The rectangular diffraction grating is shown in **Figure 3**. The $N$ slits are arranged with an interval $d$. It is assumed that a plane wave approaches the grating along the y-axis. The amplitude of the electron field at P is given by:

$$U(P) = -\frac{iA}{\lambda s_0} \sum_n \iint_A e^{-ik((nd+x)p+qz)} d\xi d\eta = U^{(0)}(p) \sum_{n=0}^{N-1} e^{-ikndp}$$
$$= U^{(0)}(p) \sum_{n=0}^{N-1} \exp(-ikndp) = U^{(0)} \frac{1-e^{-ikNdp}}{1-e^{-ikdp}}$$
[1]

Here,

$$U^{(0)}(P) = -\frac{2iAe^{-iks_0}}{\lambda s_0} \frac{\sin kpa}{kpa} \frac{\sin kqb}{kqb}$$
[2]

which is the amplitude of the wave through one slit (rectangle). $A$ is the constant coefficient and $k$ is an integer. $s_0$ represents the distance between the observation point and the center of the rectangular slit (**Fig. 3**).

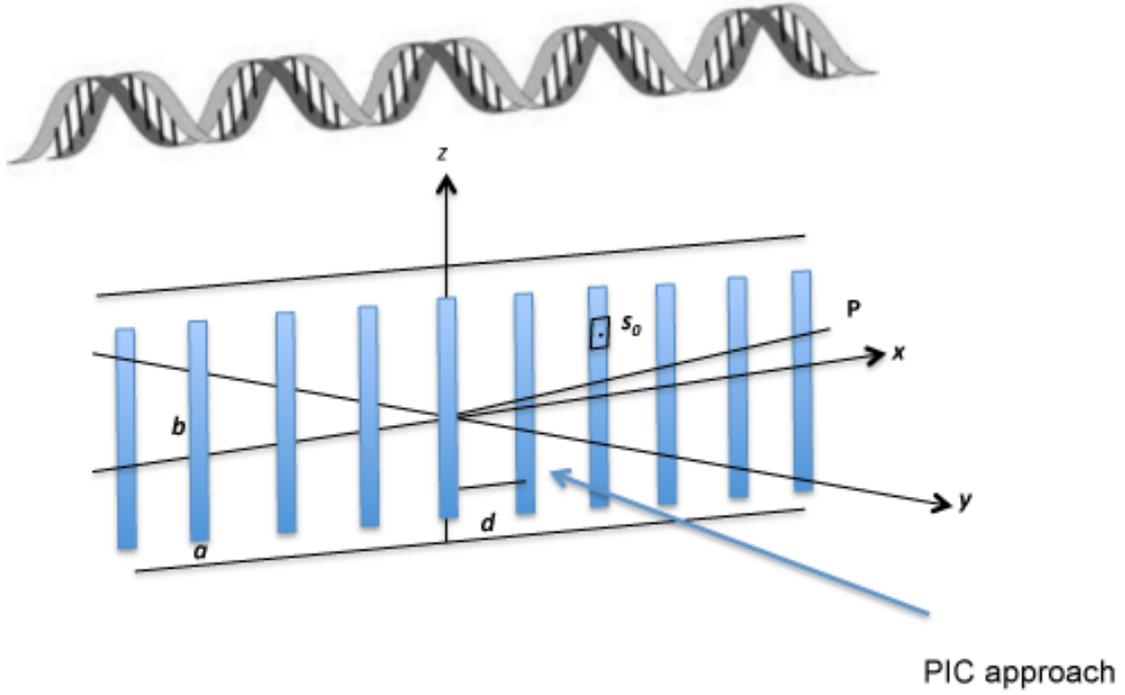

**Fig. 3. Schematic diagram of a ten-slit grating**. *a* and *b* represent the slit size. Ten rectangular slits are arranged; *P* shows the observation point and *d* represents the distance between individual slits.

Here,

$$p = x/s_0, q = z/s_0 \qquad [3]$$

In addition, the terms that depend on *p* and *q* in this equation indicate the dependence on the propagation direction of the wave. The wave intensity *I(P)* at *P* through one slit is given by the following equation:

$$I^{(0)}(P) \propto \left(\frac{AS}{\lambda s_0}\right)^2 \left(\frac{\sin kpa}{kpa}\right)^2 \left(\frac{\sin kqb}{kqb}\right)^2 \qquad [4]$$

Next, we consider that *N* rectangles are arranged at equal intervals, *d*. It is assumed that a plane moves along the *y*-axis. The center position of the $n^{th}$ rectangle is given by (*nd*, 0).

Accordingly, the light intensity is given by:

$$I(P) \propto |U(p)|^2 = \frac{1-\cos Nkdp}{1-\cos kdp} I^{(0)}(P) = \left(\frac{\sin\frac{Nkdp}{2}}{\sin\frac{kdp}{2}}\right)^2 \left(\frac{\sin kpa}{kpa}\right)^2 \left(\frac{\sin kqb}{kqb}\right)^2 \quad [5]$$

where $I^{(0)}(p)$ is equivalent to the square of the absolute value of the right side of Eq.[5], showing the intensity of the incident light that passes through a single slit. Here, $I(\Delta x)$ signifies the prospective number of mice that harbor the inserted viral DNA at the $\Delta x$-nucleotide position from the cytosine (position **r**) at which the insertion occurred most frequently. As a result, the muber of the mice with individual sites is plotted in **Fig. 1b and 2** by dotted lines. The maximum value of the light intensity is set to 100. The curve was calculated as follows:

$$I(P) \propto \left(\frac{\sin\frac{50\Delta x}{2}}{\sin\frac{5\Delta x}{2}}\right) \left(\frac{\sin \Delta x}{\Delta x}\right)^2 \quad [6]$$

Here, we set $kpa = \Delta x$ , $d = 5a$, $N = 10$, in Eq. [5] This slit number $N$, ten, was interestingly coincident with the number of nucleotides present in one full twist of a double helix, or the number of both major and minor grooves within the 50-bp segment. The slit interval $d$ (= $5a$) was coincident with the number of nucleotide pairs between the minor and major grooves, and was also very similar to the number of nucleotides recognized by retroviral DNA-ends in the PIC. These interesting coincidences suggest that the DNA structure contributes to the distribution of insertion sites (**Fig. 1b**, **2**). To adjust the amplitude of the periodic function to the actual number of mice with individual insertional mutation in *Zfp521*, the coefficient was set to 0.35 for the calculation. By setting the proportional coefficient in Eq. [6] equal to 0.35, we could normalize the ratio of mice harboring the inserted retroviral genome at the most frequent site (position **r**). Finally, we obtained the diffraction plot derived from the formula described by Fresnel–Kirchhoff theory as follows (23):

$$I = 0.35 \left( \frac{\sin(\pi \Delta x)}{\sin(\pi \Delta x / 10)} \right)^2 \left( \frac{\sin(\pi \Delta x / 25)}{\pi \Delta x / 25} \right)^2 \quad [7]$$

Here, *I(Δx)* signifies the prospective number of mice that harbor the inserted viral DNA at the *Δx*-nucleotide position from the cytosine (position **r**) at which the insertion occurred most frequently. For example, when *Δx* = 10 (bp), i.e., the **q** and **s** positions in the segment, *I(Δx)* was close to 21, which is the actual number of mice that harbored the somatically integrated MLV genome. The probability function for *N-myc* insertions is obtained by setting $kpa = \Delta x - 28$, $d = 5a$, $N = 10$ in Eq. [5] and amplitude = 0.28, instead of 0.35 in Eq.[6] (**Fig. 2**).

**Discussion**

The structure of host DNA greatly influences the site and frequency of integration. The distortion of the double helix caused by associations with nucleosomal proteins promotes the integration reaction and alters the distribution of insertion sites (8, 24). We hypothesized that (i) the PIC initially targets palindromic motifs and (ii) subsequently, the PIC actually selects the insertion sites during association and dissociation with the host DNA according to the probability function.

The collision between the PIC and host DNA occurs prior to nick introduction to the host DNA. Host DNA may initially be attacked at the middle position of the alternating palindrome or mirror symmetry by PIC; afterwards, the position or the neighboring position may be selected. The 10-bp periodicity may be linked to the nucleosome structure. However, if the periodicity uniquely depends on the 10-bp periodical structure, the frequency of insertions at the **p, q, r, s,** and **t** sites in **Fig. 1b** would be equivalent; hence, it would be difficult to explain differences in insertion frequencies among these five insertion sites and the symmetry of frequencies with respect to the **r** site. One explanation for the periodic insertions is the rewinding of the host DNA during the attachment of the PIC to the host DNA. If attachment promotes the rewinding of the segment around the hotspot **r,** the periodic open structure is anticipated around the hotspot. However, because integrase lacks topoisomerase activity, this explanation is controversial. Thus, the current data indicate that other factors determine the insertion frequency, beyond the nucleosome structure. At present, we believe that it may be advantageous for the PIC to access the **r** position in the middle of the palindrome. Importantly, a common probability function correctly predicts the insertion sites in two different genes, implying that palindromic motifs are essential for insertion selectivity, but the nucleotides themselves are not critical for the preference.

There remains the important question of why our probability function describes the insertion frequency. Insertions are a consequence of interactions between retroviral DNA and host DNA. The released retroviral DNA may attach to the neighboring site of the prominently favored site during subsequent interactions with the host DNA. The probability function describing the frequency of these events can be obtained from a detailed analysis of stochastic processes or a quantum mechanics framework. Ardnt et al. suggested a wave–particle duality of macromolecules, even viral particles (16). The MLV and host DNA involved in the insertion process had the same molecular weight order of $C_{60}$, and the insertion selectivity may be partially explained by a similar pattern of diffraction to that of $C_{60}$.

Further, the model presented here provides general insights into gene mutation; it is not possible to predict mutation sites with complete accuracy, but in the current study, the probability of insertion sites was effectively predicted by a function. The SL/Kh mice spontaneously developed MLV genome insertions in *Zfp521* or *N-myc*, without exogenous infection by MLV. To date, the prediction of nucleotide mutations in many susceptible genes has been impossible. Because an insertion in *Zfp521* was observed in nearly all SL/Kh mice, the precise nucleotide within the gene that is targeted for mutation can be predicted for an individual mouse at birth. In this sense, the current model explains the fate of nucleotide mutations at the level of the whole genome. In conclusion, we described a property of genetic DNA that determines the site of DNA mutation.

**Acknowledgments**

This work was supported by a Grant-in-Aid from the Ministry of Education, Culture, Sports, Science, and Technology, Japan (Project No. 17013086; http://kaken.nii.ac.jp/ja/p/17013086).

**References**

1. Warming, S., P. Liu, T. Suzuki, K. Akagi, S. Lindtner, G. N. Pavlakis, N. A. Jenkins, and N. G. Copeland. 2003. Evi3, a common retroviral integration site in murine B-cell lymphoma, encodes an EBFAZ-related Kruppel-like zinc finger protein. Blood 101:1934-1940.
2. Holman, A. G., and J. M. Coffin. 2005. Symmetrical base preferences surrounding HIV-1, avian sarcoma/leukosis virus, and murine leukemia virus integration sites. Proc Natl Acad Sci U S A 102:6103-6107.
3. Lewinski, M. K., M. Yamashita, M. Emerman, A. Ciuffi, H. Marshall, G. Crawford, F. Collins, P. Shinn, J. Leipzig, S. Hannenhalli, C. C. Berry, J. R. Ecker, and F. D. Bushman. 2006. Retroviral DNA integration: viral and cellular determinants of target-site selection. PLoS Pathog 2:e60.
4. Pruss, D., R. Reeves, F. D. Bushman, and A. P. Wolffe. 1994. The influence of DNA and nucleosome structure on integration events directed by HIV integrase. J Biol Chem 269:25031-25041.
5. Schroder, A. R., P. Shinn, H. Chen, C. Berry, J. R. Ecker, and F. Bushman. 2002. HIV-1 integration in the human genome favors active genes and local hotspots. Cell 110:521-529.
6. Tsuruyama, T., T. Hiratsuka, G. Jin, Y. Imai, H. Takeuchi, Y. Maruyama, K. Kanaya, M. Ozeki, T. Takakuwa, H. Haga, K. Tamaki, and T. Nakamura. 2011. Murine leukemia retrovirus integration induces the formation of transcription factor complexes on palindromic sequences in the signal transducer and activator of transcription factor 5a gene during the development of pre-B lymphomagenesis. Am J Pathol 178:1374-1386.
7. Wu, X. L., Y. Li, B. Crise, and S. M. Burgess. 2003. Transcription start regions in the human genome are favored targets for MLV integration. Science 300:1749-1751.
8. Wang, G. P., A. Ciuffi, J. Leipzig, C. C. Berry, and F. D. Bushman.

2007. HIV integration site selection: analysis by massively parallel pyrosequencing reveals association with epigenetic modifications. Genome Res 17:1186-1194.
9. Lenz, J., R. Crowther, A. Straceski, and W. Haseltine. 1982. Nucleotide sequence of the Akv env gene. J Virol 42:519-529.
10. Correa, D., E. Hesse, D. Seriwatanachai, R. Kiviranta, H. Saito, K. Yamana, L. Neff, A. Atfi, L. Coillard, D. Sitara, Y. Maeda, S. Warming, N. A. Jenkins, N. G. Copeland, W. C. Horne, B. Lanske, and R. Baron. 2010. Zfp521 is a target gene and key effector of parathyroid hormone-related peptide signaling in growth plate chondrocytes. Dev Cell 19:533-546.
11. Honda, H., N. Yamasaki, K. Miyazaki, A. Nagamachi, R. Koller, H. Oda, M. Miyazaki, T. Sasaki, Z. Honda, L. Wolff, and T. Inaba. 2010. Identification of Zfp521/ZNF521 as a cooperative gene for E2A-HLF to develop acute B-lineage leukemia. Oncogene 29:1963-1975.
12. Hiai, H., T. Tsuruyama, and Y. Yamada. 2003. Pre-B lymphomas in SL/Kh mice: a multifactorial disease model. Cancer Sci 94:847-850.
13. Tsuruyama, T., T. Hiratsuka, and N. Yamada. 2016. Hotspots of MLV integration in the hematopoietic tumor genome. Oncogene.
14. Hiratsuka, T., Y. Takei, R. Ohmori, Y. Imai, M. Ozeki, K. Tamaki, H. Haga, T. Nakamura, and T. Tsuruyama. 2015. ZFP521 contributes to pre-B-cell lymphomagenesis through modulation of the pre-B-cell receptor signaling pathway. Oncogene.
15. DePinho, R. A., E. Legouy, L. B. Feldman, N. E. Kohl, G. D. Yancopoulos, and F. W. Alt. 1986. Structure and expression of the murine N-myc gene. Proc Natl Acad Sci U S A 83:1827-1831.
16. Setoguchi, M., Y. Higuchi, S. Yoshida, N. Nasu, Y. Miyazaki, S. Akizuki, and S. Yamamoto. 1989. Insertional activation of N-myc by endogenous Moloney-like murine retrovirus sequences in macrophage cell lines derived from myeloma cell line-macrophage hybrids. Mol Cell Biol 9:4515-4522.
17. Nielsen, A. A., A. B. Sorensen, J. Schmidt, and F. S. Pedersen. 2005. Analysis of wild-type and mutant SL3-3 murine leukemia virus

insertions in the c-myc promoter during lymphomagenesis reveals target site hot spots, virus-dependent patterns, and frequent error-prone gap repair. J Virol 79:67-78.

18. Hwang, H. C., C. P. Martins, Y. Bronkhorst, E. Randel, A. Berns, M. Fero, and B. E. Clurman. 2002. Identification of oncogenes collaborating with p27Kip1 loss by insertional mutagenesis and high-throughput insertion site analysis. Proc Natl Acad Sci U S A 99:11293-11298.

19. Arndt, M., O. Nairz, J. Vos-Andreae, C. Keller, G. van der Zouw, and A. Zeilinger. 1999. Wave-particle duality of C(60) molecules. Nature 401:680-682.

20. Gatti, A., E. Brambilla, and L. A. Lugiato. 2003. Entangled imaging and wave-particle duality: from the microscopic to the macroscopic realm. Phys Rev Lett 90:133603.

21. Di Stefano, O., A. Ridolfo, S. Portolan, and S. Savasta. 2011. Test of the all-optical control of wave-particle duality of cavity photons by ordinary photodetection. Opt Lett 36:4509-4511.

22. Langham, J., and D. Barkley. 2013. Non-specular reflections in a macroscopic system with wave-particle duality: spiral waves in bounded media. Chaos 23:013134.

23. Gil, C., I. Chaib-Oukadour, and J. Aguilera. 2003. C-terminal fragment of tetanus toxin heavy chain activates Akt and MEK/ERK signalling pathways in a Trk receptor-dependent manner in cultured cortical neurons. Biochemical Journal 373:613-620.

24. Bowerman, B., P. O. Brown, J. M. Bishop, and H. E. Varmus. 1989. A nucleoprotein complex mediates the integration of retroviral DNA. Genes Dev 3:469-478.

**Supporting Material**

**Supplementary Text**

**Derivation of the probability wave function predicting mutation sites by Fresnel–Kirchhoff theory for the diffraction of hypothesized wave**

The wave function for *Zfp521* insertion is obtained by the following mathematical cords:

datapts =

 Table[{{1, 1}, {2, 2}, {3, 1}, {4, 4}, {5, 5}, {6, 4}, {7, 1}, {8, 2},{9, 2}, {10, 1}, {11, 1}, {12, 2}, {13, 2}, {14, 2}, {15, 21}, {16, 1}, {17, 2}, {18, 1}, {19, 1}, {20, 1}, {21, 1}, {22, 2}, {23, 1}, {24, 2}, {25, 35}, {26, 1}, {27, 1}, {28, 1}, {29, 2}, {30, 1}, {31, 2}, {32, 2}, {33, 1}, {34, 5}, {35, 19}, {36, 1}, {37, 2}, {38, 1},{39, 2}, {40, 4}, {41, 3}, {42, 1}, {43, 1}, {44, 11}, {45, 3}, {46, 1}, {47, 1}, {48, 0}, {49, 4}, {50, 1}, {51, 1}, {52, 0}, {53, 0}, {54, 0}, {55, 0}}]

p1 = ListLinePlot[datapts, PlotRange -> All, PlotStyle -> {Thickness[0.005], RGBColor[1, 0, 0], Dashing[{0.01, 0.01}]}]

p2 = Plot[0.35 (10*Sin[Pi/25*(x - 25)]/(Pi/25*(x - 25)))^2, {x, 0, 25 - 10^(-7)}, PlotRange -> All]

p3 = Plot[0.35 (Sin[Pi/25*(x - 25)]/(Pi/25*(x -25)))^2*(Sin[10*5*(Pi/25*(x - 25))/2]/Sin[5*(Pi/25*(x - 25))/2])^2, {x, 25 + 10^(-7), 50}, PlotRange -> All]

Show[p1, p2, p3]

'datapts' in the first cord shows the number of insertions that habour the insertion at individual nucleotides. In the notation {*i*, *j*}, '*i*' signifies the nucleotide number and '*j*' signifies the number of insertions.

**MLV insertion into *N-myc*.** The wave function for *N-myc* insertion is obtained by setting $kpa = \Delta x - 28$, $d = 5a$, $N = 10$ in (1.7)(**Figure 2**). To adjust the amplitude to the actual number of mice used, the coefficient was set to 0.08. The mathematical cords are given by:

datapts =

Table[{{1, 0}, {2, 0}, {3, 0}, {4, 0}, {5, 0}, {6, 0}, {7, 1}, {8, 0},{9, 0}, {10, 0}, {11, 0}, {12, 0}, {13, 0}, {14, 0}, {15, 1}, {16, 1}, {17,3}, {18, 1}, {19, 0}, {20, 0}, {21, 0}, {22, 0}, {23, 0}, {24, 0}, {25, 0}, {26, 0}, {27, 0}, {28, 8}, {29, 0}, {30, 1}, {31, 0}, {32, 1}, {33, 0}, {34, 0}, {35, 0}, {36, 0}, {37, 0}, {38, 4},{39, 0}, {40, 0}, {41, 0}, {42, 1}, {43, 0}, {44, 1}, {45, 0}, {46, 1}, {47, 0}, {48, 2}, {49, 0}, {50, 0}, {51, 0}, {52, 0}, {53, 0}, {54, 0}, {55, 0}}]

p1 = ListLinePlot[datapts, PlotRange -> All, PlotStyle -> {Thickness[0.005], GrayLevel[0.5], Dashing[{0.01, 0.01}]}]

p2 = Plot[0.08 (Sin[Pi/25*(x- 28)]/(Pi/25*(x - 28)))^2*(Sin[10*5*(Pi/25*(x - 28))/2]/Sin[5*(Pi/25*(x - 28))/2])^2, {x, 0, 25 - 10^(-7)}, PlotRange -> All]

p3 = Plot[0.08 (10*Sin[Pi/25*(x - 28)]/(Pi/25*(x - 28)))^2, {x, 0, 25 - 10^(-7)}, PlotRange -> All]

p4 = Plot[0.08 (Sin[Pi/25*(x-28)]/(Pi/25*(x- 28)))^2*(Sin[10*5*(Pi/25*(x-28))/2]/Sin[5*(Pi/25*(x - 28))/2])^2, {x,25 + 10^(-7), 50}, PlotRange -> All]

```
p5 = Plot[0.08   (10*Sin[Pi/25*(x - 28)]/(Pi/25*(x - 28)))^2, {x, 25 + 10^(-7), 50},
PlotRange -> All]Show[p1, p2, p3, p4, p5]
```